\documentstyle[epsf]{ioplppt}
\jl{6}

\newcommand{\msolar}{M_{\odot}}

\newcommand{\mptr}{{\cal M}}
\newcommand{\fracparen}[2]{\left(\frac{#1}{#2}\right)}
\newcommand{\qddd}{\mbox{$\stackrel{\ldots}{\mbox{Q}}$}}

\newcommand{\oderiv}[2]{\frac{d #1}{d #2}}
\newcommand{\oderivn}[3]{\frac{d^{#3}\!#1}{d #2^{#3}}}

\newcommand{\oderivnf}[3]{{d^{#3}\!#1}\!/{d #2^{#3}}}

\newcommand{\E}[1]{\times 10^{#1}}
\newcommand{\Eqref}[1]{Equation~(\ref{#1})}
\newcommand{\Figref}[1]{Fig.~\ref{#1}}

\newcommand{\hide}[1]{}\newcommand{\units}{\rm\;}

\newcommand{\Cardiff}{Department of Physics and Astronomy\\University of Wales College of Cardiff, Cardiff, U.K.}
\newcommand{\AEI}{Max Planck Institute for Gravitational Physics\\The Albert Einstein Institute\\Potsdam, Germany}

\begin{document}
\title{Gravitational wave astronomy}
\author{B F Schutz}
\address{\AEI \\and \\ \Cardiff}

\begin{abstract}
The first decade of the new millenium should see the first direct detections of gravitational waves.  This will be a milestone for fundamental physics and it will open the new observational science of gravitational wave astronomy.  But gravitational waves already play an important role in the modeling of astrophysical systems.  I review here the present state of gravitational radiation theory in relativity and astrophysics, and I then look at the development of detector sensitivity over the next decade, both on the ground (such as LIGO) and in space (LISA).  I review the sources of gravitational waves that are likely to play an important role in observations by first- and second-generation interferometers, including the astrophysical information that will come from these observations.  The review covers some 10 decades of gravitational wave frequency, from the high-frequency normal modes of neutron stars down to the lowest frequencies observable from space. The discussion of sources includes recent developments regarding binary black holes, spinning neutron stars, and the stochastic background.
\end{abstract}
\pacs{95.30.Sf,95.55.Ym,04.30.Db,95.85.Sz,04.80.Nn}

\section{Introduction}

Gravity is the dominant force in most astronomical systems.  The big surprize of the last third of the 20th century was that relativistic gravity is important in so many of them.  Neutron stars represent up to 1\% (by number) of the entire stellar population of any galaxy; stellar-mass black holes seem to be abundant; massive black holes with masses between $10^6\msolar$ and several times $10^9\msolar$ seem to inhabit the centres of most galaxies; quasars, X-ray binaries, supernovae and gamma-ray bursts use relativistic gravity to convert mass into energy with efficiencies ten times or more greater than nuclear reactions can achieve; gravitational lensing has become an important tool for understanding the mass distribution of the universe; and of course the Big Bang is the only naked singularity we expect to be able to see.  Most of these systems are either dynamical or were formed in catastrophic events; many are or were therefore strong sources of gravitational radiation.  As the 21st century opens, we are on the threshold of using this radiation to gain a new perspective on the observable universe.

I have chosen the term ``gravitational wave astronomy'' as the title for this review deliberately.  It suggests, of course, a branch of observational astronomy.  Since it may still be 10 or more years before this observational tool becomes established and reliable, it might be thought that my use of this term is premature.  However, gravitational wave astronomy already exists.  The theory of gravitational radiation already makes an important contribution to the understanding of a number of astronomical systems, such as binary neutron stars, cataclysmic variables, young neutron stars, low-mass X-ray binaries, and even the anisotropy of the microwave background radiation. As the understanding of relativistic systems improves, it can be expected that gravitational radiation will become more and more important as a theoretical tool.

Naturally, the most exciting prospect for the field is the direct observation of gravitational waves. Not only could sufficiently detailed observations verify the predictions of general relativity, in particular about their polarization, but they could also potentially reveal the existence of a weak scalar component of gravity, which may be expected from many kinds of unified field theories.  Moreover, direct observations would reveal information about astronomical systems that is available in no other way.
\begin{itemize}
\item Because gravitational waves are emitted by the bulk motions of their sources, not by individual atoms or electrons, as is normally the case for electromagnetic waves, they carry a completely different kind of information about their sources from that which is normally available from other observations.  For example, the polarization of waves from the orbit of a binary system reveals the inclination of the orbit to the line of sight, a crucial unknown in the modelling of many such systems.
\item Gravitational waves provide the only way to make direct observations of black holes.  It is the only radiation they can emit with observable strength.  All other information about black holes is indirect, coming from their effects on gas in their environments.
\item Gravitational waves interact with matter so weakly that they are not attenuated or scattered on their way to the detector.  (They can be gravitationally lensed, however.)  This means that they can reveal information about hidden regions, such as the interior of a supernova explosion or the Big Bang. 
\end{itemize}

The first three-quarters of the 20th century were required to place the mathematical theory of gravitational radiation on a sound footing. Many of the most fundamental constructs in general relativity, such as null infinity and the theory of conserved quantities, were developed at least in part to help solve the technical problems of gravitational radiation. I will not cover this history here, for which there are excellent reviews.\cite{MTW, DAMOUR1987} There are still many open questions, since it is impossible to construct exact solutions for most interesting situations.  Among the most important difficulty is that we lack a full understanding of the two-body problem, and I will review the theoretical work on this problem below.  But the fundamentals of the theory of gravitational radiation are no longer in doubt.

This review is divided into three main parts, plus the introduction and conclusions.  The first part, \ref{sec:current}, treats the existing theoretical work on gravitational radiation.  I first discuss the present activity in gravitational wave astronomy, and I follow that by describing theoretical and numerical work on gravitational radiation emitted by orbiting bodies: the relativistic two-body problem.  The second part, \ref{sec:detectors}, covers the development of gravitational wave detectors.  There is a brief history  followed by a description of the principles underlying the operation of both resonant-mass and beam detectors.  Planned detector developments, including a detector in space, are also reviewed.  The final part of the review, \ref{sec:sources}, treats the expected sources of gravitational radiation that are likely to be detectable in the next decade or so.

The scope of this review is too wide to allow a treatment of any of these topics in depth.  I will try, however, to describe where things stand at present in these fields, sometimes with a perspective on their historical development, and I will give references where the interested reader can find further information.

\section{Gravitational wave theory}\label{sec:current}
\subsection{Gravitational wave astrophysics today}\label{sec:gw_astro}
Gravitational radiation plays an observable role in the dynamics of many known astronomical systems.  In some, such as cataclysmic variables and binary neutron star systems, the role of gravitational radiation has been understood for years.  In others, such as young neutron stars and low-mass X-ray binaries, the importance of gravitational radiation has been understood only recently.  As further observations, particularly at X-ray wavelengths, become available, the importance of gravitational radiation as a tool for modelling astronomical  systems should increase.

\subsubsection{Cataclysmic variables.} The first example of the use of gravitational radiation in modelling an observed astronomical system was the explanation by Faulkner \cite{Faulkner1971} of how the activity of cataclysmic binary systems is regulated.  Such systems, which include many novae, involve accretion by a white dwarf from a companion star. Unlike accretion onto neutron stars, where the accreted hydrogen is normally processed quickly into heavier elements, on a white dwarf the unprocessed material can build up until there is a nuclear chain reaction, which results in an outburst of visible radiation from the system.   

Now, in a circular binary system that conserves total mass and angular momentum, a transfer of mass from a more massive to a less massive star will make the orbit shrink, while a transfer in the opposite direction makes the orbit grow.  If accretion onto a white dwarf begins with the dwarf as the less massive star, then the stars will draw together, and the accretion will get stronger.  This runaway stops when the stars are of equal mass, and then accretion begins to drive them apart again. Astronomers observed that in this phase accretion in certain very close binaries continued at a more or less steady rate, instead of shutting off as the stars separated more and more.  Faulkner pointed out that gravitational radiation from the orbital motion would carry away angular momentum and drive the stars together.  The two effects together result in steady accretion at a rate that can be predicted from the quadrupole formula and simple Newtonian orbital dynamics, and which is in good accord with observations. 

\subsubsection{Binary neutron stars.} The most famous example of the effects of gravitational radiation on an orbiting system is the Hulse-Taylor Binary Pulsar, PSR1913+16.  In this system, two neutron stars orbit in a close eccentric orbit.  The pulsar provides a regular clock that allows one to deduce, from post-Newtonian effects, all the relevant orbital parameters and the masses of the stars.  The quadrupole formula of general relativity \cite{MTW} then predicts the orbital shrinking without any adjustable parameters, and the result is in accord with observations within the observational errors, which are below 1\% \cite{DAMOUR1992a}.  This is the most sensitive test that we have of the correctness of Einstein's equations in respect of gravitational radiation, and it leaves little room for doubt in the validity of the quadrupole formula for other systems that may generate detectable radiation.

\subsubsection{Young neutron stars --- the r-mode instability.} In 1971 Chandrasekhar \cite{ChandraCFS} applied the quadrupole formula to calculate the corrections to the eignefrequencies of the normal mode vibrations of rotating stars, and he found to his surprise that some modes were made unstable, i.e. that coupling to gravitational radiation could destabilize a rotating star.  Subsequent work by Friedman and Schutz \cite{Friedman1978} showed that there was a key signature for a mode of a Newtonian star that would be unstable in general relativity.  This was the pattern speed of the mode, i.e. the angular velocity at which the crests of the pattern rotated about the rotation axis of the star. If this speed was in the same sense as the rotation of the star, but slower than the star, then the mode would be unstable in a perfect-fluid star. This instability has come to be known as the CFS instability, after the three authors who explained it.  

The basic theory was developed for perfect-fluid stars.  However, Lindblom and Detweiler \cite{LINDBLOM1977} showed that the effect of viscosity ran counter to that of radiation reaction, so that the instability was strongest in modes with the longest wavelengths, i.e. in the quadrupolar modes.  Full numerical calculations on Newtonian stellar models with realistic viscosity models showed \cite{Lindblom1995} that the standard fundamental and acoustic modes of rotating neutron stars were not vulnerable to this instability.  Subsequent work on fully relativistic models \cite{Stergioulas1998} has hinted that the instability may be stronger than the Newtonian models indicate, but it is still at the margins of astrophysical interest. T

The situation changed in 1997 when Andersson \cite{Andersson1998} pointed out that there is another class of modes of Newtonian stars that should be unstable in the same way, but which had not been studied in this context before, the so-called Rossby or r-modes.  These are momentum-dominated modes, where the gravitational radiation comes from the current-quadrupole terms rather than from the mass quadrupole.  Investigations by a number of authors \cite{Lindblom1998,Andersson1999,Owen1999} have shown that this instability is very strong in hot, rapidly rotating stars.  This is particularly relevant to young neutron stars, which may well be formed with rapid spin and which will certainly be hot.  For their first year, stars spinning faster than about 100~Hz will spin down to about 100~Hz by losing angular momentum to gravitational radiation.  This radiation may be detectable by future detectors, and may also form a strong cosmic background at frequencies above 20~Hz.  But the implication of most immediate importance is that it explains why all known young neutron stars are relatively slow rotators, rotating ten times slower than the (older) millisecond pulsars.  Work over the next few years should considerably refine our picture of the early evolution of young neutron stars.

\subsubsection{Low-mass X-ray binaries.} Recent observations by the Rossi satellite (RXTE) have given evidence that the class of X-ray sources called Low-Mass X-ray Binaries (LMXB's) contains neutron stars with a remarkable narrow range of spins, between perhaps 250~Hz and 320~Hz.\cite{vanderKlis1998}  These are systems in which it is believed that neutron stars are spun up from the low angular velocities they have after their lifetime as normal pulsars to the high spins that millisecond pulsars have.  One would expect, therefore, that the spins of neutron stars in such systems would be spread over a wide range. The fact that they are not requires an explanation.

The most viable explanation so far offered is the suggestion of Bildsten \cite{Bildsten1998} that gravitational radiation limits the rotation rate.  The proposed mechanism is that anisotropic accretion onto the star creates a temperature gradient in the crust of the neutron star, which in turn creates a gradient in the mass of the nucleus that is in local equilibrium, and this in turn creates a density gradient that leads, via the rotation of the star, to the emission of gravitational radiation.  This radiation carries away angular momentum, balancing that which is accreted, so that the star remains at an approximately constant speed.

This mechanism will receive further study, and observations with new satellites will test the model stringently.  According to the model, the gravitational wave luminosity of the star is proportional to the measured flux of X-rays, since the X-ray flux is itself proportional to the accreted angular momentum that has to be carried away by the gravitational waves.  If this model is correct, then the X-ray source Sco X-1 will be marginally detectable by the first generation of interferometers now under construction.  

\subsubsection{Cosmic background radiation and galaxy formation.} The initial observations of the anisotropy of the cosmic microwave background by the COBE satellite \cite{SMOOT1992} opened a new window on the early universe.  The next generation of satellites, called MAP (to be launched by NASA) and PLANCK (by ESA), will enormously improve the sensitivity to small-scale anisotropies.  The observations at present do not discriminate between anisotropies produced by density perturbations and those produced by gravitational radiation at the epoch of recombination.  The gravitational wave content is critical to understanding early galaxy formation and using the anisotropies to measure cosmological parameters.  Gravitational wave detectors can in principle measure the gravitational wave background, but only at much shorter wavelengths than those that affect satellite measurements.\cite{Allen1997}  Indeed, there may be no close relation between a measured background and the radiation at long wavelengths \cite{Grishchuk1997,Buonanno1997}.  But the upcoming satellite observations may tell us how much is gravitational radiation and how much is density perturbation.  We will return to this question below.

\subsection{The two-body problem}\label{sec:gw_theory}
The largest effort in gravitational radiation theory in recent years has been to study the two-body problem using various approximations.  The reason is that binary systems are likely to be important gravitational wave sources, and until the evolution of such a system is thoroughly understood, it will not be possible to extract the maximum information from the observations.  

The two-body problem must be solved approximately because of the two difficulties of handling the radiation field and of the nonlinearity of Einstein's equations.  One way to approach this problem is to solve Einstein's equations numerically.  Another is to employ an analytic approximation scheme. Most such schemes make use of expansions in the smallness of either the velocity $v$ (alternatively the inverse speed of light $1/c$) and/or the internal Newtonian potential $\phi_{int}\sim M/R$ (for a source of mass $M$ and size $R$). In the most important approximation scheme, the post-Newtonian scheme, these two parameters are linked because $v^2\sim M/R$ for self-gravitating systems, even relativistic ones.  

\subsubsection{Numerical approaches to the two-body problem.}  From the point of view of relativity, the simplest two-body problem is that of two black holes.  There are no matter fields and no point particles, just pure gravity.  A number of teams are working towards developing accurate numerical solutions for the coalescence of two black holes from orbit, using fully three-dimensional numerical simulations.  

Progress has been steady, but the task is enormous.  The main impediments are algorithmic, in that it has so far proved difficult to find numerical schemes that handle the outer boundary of the numerical grid in a satisfactory way (allowing waves to leave without reflection) and to impose inner boundaries around the black-hole singularities that move with the black holes in a stable manner.  Significant progress on the second problem was recently described by the Binary Black Hole Grand Challenge Alliance \cite{Cook1998}, but robust schemes that can handle sufficiently long integration times are still lacking.  

The most general black-hole collision yet simulated was performed only recently by the NCSA/Potsdam/Wash-U collaboration \cite{NCSA1999}.  Two black holes were simulated from a starting position very near to one another, but their initial velocities were not along the line joining them, and their initial masses and spins were very different.  The simulation could follow the emission of gravitational radiation and the formation of a single apparent horizon around both, but since it was done in a coordinate system that did not cut out the singularities, the simulation had a limited duration in time.  Although this represents a big advance in state-of-the-art simulations, it is still far from what is needed for the interpretation of gravitational wave observations.

\subsubsection{Analytic approximations to the two-body problem.} For the interpretation of observations of binary neutron-star coalescences, which might be detected within 5 years by detectors now under construction, it is necessary to understand their orbital evolution to a high order in the post-Newtonian expansion.  The first effects of radiation reaction are seen at 2.5-pN order, but we probalby have to have control over the expansion at least to 3.5-pN order.  There are many approaches to this, and I can not do justice here to the enormous effort that has gone into this field in recent years.

The most successful methods so far have come from treating a binary system as if it were composed of two point masses.  This is, strictly speaking, inconsistent in general relativity, since the masses should form black holes of finite sizes. Blanchet, Damour, and collaborators \cite{Blanchet} have avoided this problem by a method that involves generalized functions.  They first expand in the nonlinearity parameter, and when they have reached sufficiently high order they obtain the velocity expansion of each order.  By ordering terms in the post-Newtonian manner they have developed step-by-step the approximations up to 3-pN order.  A different team, led by Will, works with a different method of regularizing the point-particle singularity and compares its results with those of Blanchet et al at each order \cite{Blanchet1995}.  There is no guarantee that either method can be continued successfully to any particular order, but so far they have worked well and are in agreement.  Their results form the basis of the templates that are being designed to search for binary coalescences.  An interesting way of extending the validity of the expansion that is known to any order is to apply Pad\'{e} approximants \cite{Damour1998}.

Other methods have been applied to this problem.  Futamase \cite{Futamase} introduced a limit that combines the nonlinearity and velocity expansions in different ways in different regions of space, so that the orbiting bodies themselves have a regular (finite relativistic self-gravity) limit while their orbital motion is treated in a Newtonian limit.  This should not fail at any order, but it has a degree of arbitrariness in choosing initial data that could cause problems for gravitational wave search templates that integrate orbits for a long period of time.  Linear calculations of point particles around black holes are of interest in themselves (see the LISA project below) and also for checking results of the full two-body calculations.  These are well-developed for certain situations, e.g. \cite{Tagoshi1996,Mino1999}.  But the general equation of motion for such a body, taking into account all non-geodesic effects, has not yet been cast into a form suitable for practical calculations. \cite{Capon1999,Quinn1997}

\section{Detectors for gravitational radiation}\label{sec:detectors}
\subsection{A 40-year history}\label{sec:history}
The first practical instruments for detecting gravitational radiation were constructed by Joseph Weber \cite{Weber1960}.  Based on massive cylinders of aluminium, these so-called ``bar'' detectors were the best technology at the time for gravitational wave detection.  These detectors exploit the sharp resonance of the cylinder to get their sensitivity, which is normally confined to a narrow bandwidth (one or a few Hz) around the resonant frequency.  

In the 1970's a number of groups turned to laser interferometry for the basis of a new kind of detector.  This technique was considered by Weber, but the technology available in the 1960's did not make it a good choice. Improvements in lasers and mirrors changed that picture, and by the early 1980's three prototype interferometers were operating in Glasgow \cite{DREVER1983b}, in Garching, near Munich \cite{BILLING1983}, and at MIT.\cite{BENFORD1991}  A prototype at Caltech followed soon after.\cite{DREVER1991}

While bar detectors continue to be developed, and have until very recently had a sensitivity to broadband bursts that was superior to that of the existing interferometers, the best hope for the first detections of gravitational radiation lies with large-scale interferometers that have developed out of the prototypes mentioned above. In the next decade we may also see the launch of a space-based interferometer, LISA, to search for signals at frequencies lower than those that are accessible from the ground.  

In this section I review these developments.  Two sections (\ref{sec:barprinciples} and \ref{sec:beamprinciples}) describe breifly the physical principles underlying the operation of the two main kinds of detectors, resonant masses and beam detectors (using light). I then describe the large-scale interferometers now under construction and look ahead at their long-term future (Section~\ref{sec:present}).
Finally I describe the space-based project LISA and consider the exciting prospects for the long-term future of gravitational wave detection in space.  For more on the history of the development of detectors, see the review by Thorne \cite{THORNE1987}.  An excellent introduction to interferometers is the book by Saulson.\cite{SAULSON1994}

\subsection{Principles of the operation of resonant mass detectors}\label{sec:barprinciples}
A typical ``bar'' detector consists of a cylinder of aluminium with a length $L\sim 3$~m,  a resonant frequency of order $f\sim 500$~Hz to 1.5~kHz, and a mass $M\sim 1000$~kg. A short gravitational wave burst 
with $h\sim 10^{-21}$ will make the bar 
vibrate with an amplitude 
\begin{equation}
\delta \ell_{gw} \sim h\ell \sim 10^{-21}\;\rm m. 
\end{equation}
To measure this, one must fight against three main sources of noise:
\begin{enumerate}
\item {\bf Thermal noise.} The original Weber bar operated at room temperature, 
but the most advanced detectors today, Nautilus\cite{COCCIA1995b} and Auriga,\cite{CERDONIO1995} operate at $T = 100$~mK.  At this temperature the r.m.s.\ amplitude of vibration is 
\begin{equation}
\langle \delta \ell^2 \rangle^{1/2}_{th} = \left(\frac{kT}{4 \pi^2 M f^2} \right)^{1/2} \sim 6\times 10^{-18}\;\rm m. 
\end{equation}
This is far larger than the gravitational wave amplitude.  But if the material has a high 
$Q$ (say $10^6$) in its fundamental mode, then it changes its thermal amplitude of vibration in 
a random walk with very small steps, taking a time $Q/f\sim 1000$~s to 
change by the full amount.  However, a gravitational wave burst will 
cause a change in 1~ms.  In 1~ms, thermal noise will have random-walked to 
an expected amplitude change  $(1000\rm\;s/1\;ms)^{1/2} = Q^{1/2}$ times 
smaller, or (for these numbers)
\begin{equation}
\langle \delta \ell^2 \rangle^{1/2}_{th:\;1\;\rm ms} = \left(\frac{kT}{4 \pi^2 M f^2Q} \right)^{1/2} \sim 6\times 10^{-21}\;\rm m. 
\end{equation}
So bars today can approach the goal of 
detection at or slightly below $h=10^{-20}$ against thermal noise.

\item {\bf Sensor noise.} A transducer converts the bar's mechanical 
energy into electrical energy, and 
an amplifier  increases the electrical 
signal to record it.  If sensing of the vibration could 
be done perfectly, then the detector would be broad-band: 
both thermal impulses and gravitational wave forces are mechanical forces, 
and the ratio of their induced vibrations would be the same at all 
frequencies for a given applied impulsive force.  

But sensing is not 
perfect: amplifiers introduce noise, and this makes small 
amplitudes harder to measure.  The amplitudes of vibration are largest in
the resonance band near $f$, so amplifier noise limits the detector sensitivity
to gravitational wave frequencies near $f$.  But if the noise is small, then 
the measurement bandwidth about $f$ can be much larger 
than the resonant bandwidth $f/Q$.  Today, typical measurement
bandwidths are 1~Hz, about 1000 times larger than the 
resonant bandwidths.  In the near future, 
it is hoped to extend these to 10~Hz or even 100~Hz.  
\item {\bf Quantum noise.} The zero-point vibrations of a bar with a frequency 
of 1~kHz are 
\begin{equation}
\langle \delta \ell^2 \rangle^{1/2}_{quant} = \left(\frac{\hbar}{2 \pi M f} \right)^{1/2} \sim 4 \times 10^{-21}\;\rm m. 
\end{equation}
This is comparable to the thermal 
limit over 1~ms.  So if current 
detectors improve their thermal limits, 
they will run into the quantum limit, which must be breached before a signal 
at $10^{-21}$ can be seen with such a detector.  

It is also not impossible to breach the quantum limit. The uncertainty principle 
only sets the limit above if a measurement tries to determine the excitation 
energy of the bar, or equivalently the phonon number.  But one is not interested 
in the phonon number, except in so far as it allows one to determine the 
original gravitational wave amplitude. 
It is possible to define other observables that also respond to the 
gravitational wave and which 
can be measured more accurately by {\bf squeezing} their uncertainly 
at the expense of greater errors in 
their conjugate observable.\cite{Caves1980} 
No viable schemes to do this for bars have been demonstrated so far, although 
squeezing is now an established technique in quantum optics. 
\end{enumerate}

For the foreseeable future, one can expect that bars will remain fairly narrow-band 
detectors, and that they will have difficulty getting below a sensitivity limit 
of $10^{-21}$.  These limitations motivated groups to explore the intrinsically 
wideband technique of laser interferometry.  However, resonant mass detectors have a future. It is possible now to construct large spheres of a similar size (1 to 3~m diameter) as existing cylinders. This increases the 
mass of the detector and also improves its direction-sensing. One can  in principle push to below $10^{-21}$ with spheres.\cite{COCCIA1995e} The excellent sensitivity of resonant detectors within their narrow bandwidths makes them suitable for specialized, high-frequency searches.  

\subsection{Principles of the operation of beam detectors.}\label{sec:beamprinciples}
Interferometers use laser light to measure changes in the difference between the lengths of two 
perpendicular (or nearly perpendicular) arms. Typically the arm lengths respond differently to 
a given gravitational wave, so an interferometer is a natural instrument to measure gravitational waves.  But other detectors also use electromagnetic radiation, for example ranging to spacecraft in the solar system. 

All such detectors are governed by a fundamental equation, which describes the 
interaction of light with a gravitational wave. If, for weak linearized gravitational waves, a beam of light travels a distance $L$ outwards and then back again, its round-trip time is affected by the gravitational wave. All beam detectors measure this, one way or another. If a plane gravitational wave with amplitude $h$ starts at time t and moves at an angle $\theta$ to the direction in which the beam is travelling, and if its return time $t_{return}$ is measured as proper time on a clock that remains at the point where the light started out, then this return time varies at the rate
\begin{eqnarray}
\oderiv{t_{return}}{t}&=&1 + \frac{1}{2}\left\{(1-\cos\theta)h(t+2L) - (1+\cos\theta)h(t)\right.\nonumber\\
&&\quad \left.+2\cos\theta h[t+L(1-\cos\theta)]\right\}. \label{eqn:threeterm}
\end{eqnarray}
This three-term relation is the starting point for analyzing the response of all beam detectors. 

\subsubsection{Interferometers.}
The largest detectors under construction today are the LIGO detectors,\cite{RAAB1995} two of which have arm lengths of 4~km. This is much 
smaller than the wavelength of light, so the interaction of the detector with a gravitational wave can be well approximated by the small-$L$ approximation to Equation~\ref{eqn:threeterm}, namely
\[\oderiv{t_{return}}{t} = (1-\cos\theta)^2L\dot{h}(t).\]

A detector with an arm length of 4~km responds to a gravitational wave 
with an amplitude of $10^{-21}$ with
\[\delta l_{gw}\sim hl \sim 4\E{-18}\;\rm m. \]
Light takes only about $10^{-5}$~s to go up and down one arm, much less than the 
typical period of gravitational waves of interest. Therefore, it is beneficial 
to arrange for the light to remain in an arm longer than this, say for 100 
round-trips.  This increases its effective path length by 100 and hence 
the shift in the position of a given phase of the light beam will be of 
order $10^{-16}$~m. Most interferometers keep the light in the arms for 
this length of time by setting up optical cavities in the arms with 
low-transmissivity mirrors; these are called Fabry-Perot cavities.

The main sources of noise against which a measurement must compete are:
\begin{enumerate}
\item {\bf Ground vibration.}  External mechanical vibrations must be screened out. These are a problem for bar detectors, too, but are more serious for 
interferometers, not least because interferometers 
bounce light back and forth between the mirrors, and so each reflection introduces further vibrational noise. 
Suspension/isolation systems are based on pendula. A 
pendulum is a good mechanical filter for frequencies above its natural 
frequency.  By hanging the mirrors on pendula of perhaps 0.5~m length, 
one achieves filtering below a few Hz.  Since the spectrum of ground noise 
falls at higher frequencies, this provides suitable isolation. But these 
systems can be very sophisticated; the GEO600\cite{DANZMANN1995} detector has a three-stage 
pendulum and other vibration isolation components.\cite{Plissi1998}
The most ambitious isolation system is being developed for the Virgo detector.\cite{GIAZOTTO1995}
\item {\bf Thermal noise.} Vibrations of the mirrors and of the 
suspending pendulum can mask gravitational waves. As with vibrational noise, 
this is increased by the bouncing of the light between the mirrors.
Unlike bars, interferometers measure only 
at frequencies far from the resonant frequency, where the amplitude of 
vibration is smaller. Thus, the pendulum suspensions have thermal noise at a few Hz, but measurements will be made above 20 or 30~Hz in the first detectors. Internal vibrations of the mirrors have natural frequencies of several kiloHertz. By ensuring that both kinds of oscillations have very high Q, one can confine most of the vibration energy to a small bandwidth around the resonant 
frequency, so that at the measurement frequencies the vibration amplitudes are small. This allows interferometers to operate at 
room temperature.  But mechanical $Q$s of $10^7$ or higher are required, and 
this is technically demaning.
\item {\bf Shot noise.}  The photons that are used to do interferometry 
are quantized, and so they arrive at random and make random fluctuations 
in the light intensity that can look like a gravitational wave signal. 
The more photons one uses, the smoother will be the interference signal.
As a random process, the error improves with the square-root of the number 
$N$ of photons.  Using infrared light with a wavelength 
$\lambda\sim 1\;\mu\rm m$, one can expect to measure to an accuracy of
\[\delta l_{shot} \sim \lambda/(2\pi\sqrt{N}).\]
To measure at a frequency $f$, one has to make at least $2f$ measurements 
per second, so one can accumulate photons for a time $1/2f$.  
With light power $P$, one gets $N=P/(hc/\lambda)/(2f)$ photons.  In order that $\delta 
l_{shot}$ should be below $10^{-16}$~m one needs large light power, 
far beyond the output of any continuous laser.

Light-recycling techniques overcome this problem, by using light efficiently.
An interferometer actually has two places where light leaves.  One is where 
the interference is measured. The other goes back towards the input laser. 
Normally one arranges that no light goes to the interference sensor, so that only when a gravitational wave passes does a signal register there. This means that all the light normally returns to the mirror, apart from small losses at the mirrors.  Since mirrors are of good quality, only one part in $10^3$ or less of the light is lost during a 1~ms storage time. By placing a power-recycling 
mirror in front of the laser, one can reflect this wasted light back in, 
allowing power to build up in the arms until the laser merely resupplies
the mirror losses.\cite{DREVER1983b} This can dramatically reduce
the power requirement for the laser.  The first interferometers will work 
with laser powers of 5-10~W.  This is attainable with modern laser technology.  
\item {\bf Quantum effects.} Shot noise is a quantum noise, but in addition 
there are effects like bar detectors face: zero-point vibrations of 
mirror surfaces and so on.  These are small compared to present operating 
limits of detectors, but they may become important in 5 years or so.  
Practical schemes to reduce this noise have already been demonstrated 
in principle, but they need to be improved considerably.  They can be reduced by making the mirror masses large, since the amplitude of vibration scales inversely as the square-root of the mass. 
\item {\bf Gravity gradient noise.} One noise that cannot be screened out is that due to changes in the local Newtonian gravitational field on the timescale of the measurements. A gravitational wave detector will respond to tidal forces from local sources justs as well as to gravitational waves. Environmental noise comes not only from man-made 
sources, but even more importantly from natural ones: seismic waves are 
accompanied by changes in the gravitational field, and changes in air pressure are accompanied by changes in air density.  The spectrum falls steeply with increasing frequency, so for first-generation interferometers this will not be a problem, but it may limit the performance of detectors a decade from now. And it is the primary reason that detecting gravitational waves in the low-frequency band around 1~mHz must be done in space.
\end{enumerate}

\subsubsection{Interferometers presently under construction.}\label{sec:present}
The two largest interferometer projects are LIGO\cite{RAAB1995} and VIRGO\cite{GIAZOTTO1995}.  LIGO is building three detectors on two sites. At Hanford, Washington, there will be a 4~km and a 2~km detector in the same vacuum system. At Livingston, Louisiana, there will be a single 4~km detector, oriented to be as nearly parallel to the Hanford detector as possible.  It expects to begin serious data taking in 2001 with a broadband sensitivity near $10^{-21}$.  VIRGO is building a single 3~km detector near Pisa in Italy.  It will begin taking good data in 2002 with a similar or better sensitivity.

A smaller 600-m detector, GEO600, is under construction near Hanover, Germany.\cite{DANZMANN1995} Although smaller, it has second-generation technology (primarily in its suspensions, mirror materials and interferometry), 
and it will use this to attain a sensitivity similar to the initial LIGO sensitivity.  After a few years of operation, also beginning in 2001, its technology will be transferred to LIGO and VIRGO as part of upgrades, described below.    

A yet smaller detector, TAMA300,\cite{TSUBONO1995} with 300~m arms, has recently acheived a sensitivity of about $10^{-19}$ in Japan, and improvements will take its sensitivity to below $10^{-20}$.  It is therefore at present the most sensitive interferometer ever operated, and it is competetive with the Nautilus and Auriga bar detectors. But it is seen as a development prototype, and its sensitivity will be confined to higher frequencies (above 100~Hz).  

There are plans for a detector in Australia, and a small interferometer has recently been constructed.\cite{McClelland} But it is not yet clear if a larger detector will be funded.  It would be desirable from the point of view of the physics that we can learn from observations, that detectors should be widely spaced on the Earth, so Australia and Japan are desirable places for further development.  

Designs are already in hand at the existing detectors for their first upgrades.  LIGO expects to move to LIGO II, a collaboration with the GEO groups, by 2006. VIRGO should also be upgraded by then, again using some GEO technology. We will see below that, while the first detectors could detect gravitational waves, the second generation will have a much greater assurance of success. 

Beyond that, scientists are now studying the technologies that may be needed for a further large step in sensitivity, to third-generation detectors. This may involve cooling mirrors, using ultra-massive mirrors of special materials, using purely non-transmissive optics, and even circumventing the quantum limit in interferometers, as has been studied for bars. The goal of third-generation detectors would be to be limited just by gravity-gradient noise and quantum effects. 

\subsection{Detection from space}
\subsubsection{Ranging to spacecraft.}  Both NASA and ESA perform experiments in which they monitor the return time of communication signals with interplanetary spacecraft for the characteristic effect of gravitational waves.  For missions to Jupiter and Saturn, for example, the return times are of order $2-4\E3$~s.  Any gravitational wave event shorter than this will, by Equation~\ref{eqn:threeterm}, appear 3 times in the time-delay: once when the wave passes the Earth-based transmitter, once when it passes the spacecraft, and once when it passes the Earth-based receiver. Searches use a form of data analysis using pattern matching.  Using two transmission frequencies and very stable atomic clocks, it is possible to achieve sensitivities for $h$ of order $10^{-13}$, and even $10^{-15}$ may soon be reached.\cite{ARMSTRONG1989b} 

\subsubsection{Pulsar timing.}  Many pulsars, particularly the old millisecond pulsars, are extraordinarily regular clocks, with random timing irregularities too small for the best atomic clocks to measure.  If one assumes that they emit pulses perfectly regularly, then one can use observations of timing irregularities of single pulsars to set upper limits on the background gravitational wave field.  Here the 3-term formula is replaced by a simpler two-term expression, because we only have a one-way transmission.  Moreover, the transit time of a signal to the Earth from the pulsar may be thousands of years, so we cannot look for correlations between the two terms in a given signal.  Instead, the delay is a combination of the effects of waves at the pulsar when the signal was emitted and waves at the Earth when it is received.  Observations have been used to set limits on a background of gravitational radiation at very low frequencies.\cite{Millisecond}

If one simultaneously observes two or more pulsars, the Earth-based part of the delay is correlated, and this offers a means of actually detecting long-period gravitational waves.  Observations require timescales of several years in order to achieve the long-period stability of pulse arrival times, so this method is suited to looking for strong gravitational waves with periods of several years.  Observations are currently underway at a number of observatories.

\subsubsection{Space interferometry: LISA.}\label{sec:LISA} Gravity-gradient noise on the Earth is much larger than the amplitude of any expected waves from astronomical sources at frequencies below about 1~Hz, but this noise falls off a $1/r^3$ as one moves away from the Earth.  A detector in space would not notice the Earth's noisy environment. The LISA project, currently being studied by both ESA and NASA with a view toward a collaborative mission around 2010, would open up the frequency window between 0.1~mHz and 0.1~Hz for the first time.\cite{HOUGH1995}

We will see below that there are many exciting sources expected in this waveband, for example the coalescences of giant black holes in the centres of galaxies. LISA would see such events with extraordinary sensitivity, recording typical signal-to-noise ratios of 1000 or more for events at redshift 1.  

An space-based interferometer can have arm lengths much greater than a wavelength. LISA, for example, would have arms $5\times10^6$~km long, and that would be longer than half a wavelength for any gravitational waves above 30~mHz. In this regime, the response of each arm will follow the three-term formula we encountered earlier. 

LISA would actually consist of three free-flying spacecraft, arranged in an equilateral triangular array, orbiting the Sun at 1~AU, about 20 degrees behind the Earth in its orbit.  By passing light along each of the arms, one can construct two different interferometers, so one can measure the polarization of a gravitational wave directly.  The spacecraft are too far apart to use simple mirrors to reflect light back along an arm: the reflected light would be too weak. Instead, LISA will have optical transponders: light from one spacecraft's on-board laser will be received at another, which will then send back light from its own laser locked exactly to the phase of the incoming signal. 

The main environmental disturbance to LISA are the forces from the Sun: fluctuations in solar radiation pressure and pressure from the solar wind.  To minimize these, LISA incorporates drag-free technology.  Interferometry is 
referenced to an internal proof mass that falls freely, not attached to the spacecraft. The job of the spacecraft is to shield this mass from external disturbances. It does this by sensing the position of the mass and firing its own jets to keep itself (the spacecraft) stationary relative to the proof mass. To do this, it need thrusters of very small thrust that have accurate control.  The availability of such thrusters, of the accelerometers needed to sense distrubances to the spacecraft, and of lasers capable of continuously emitting 1~W of infrared light for years, have enabled the LISA mission.

\section{Sources of gravitational radiation}\label{sec:sources}
\subsection{Estimation of gravitational wave emission}\label{sec:emission}
At this point in the progress of gravitational wave detection, the greatest emphasis in calculations of sources is on {\em prediction}: trying to anticipate what might be seen. Not only is this important in motivating the construction of detectors, but it also guides details of their design and, very importantly, the design of data analysis methods. I will return to this issue later. 

Once gravitational waves have been observed, there will undoubtedly be a shift of emphasis to include {\em interpretation}.  This will require different kinds of calculations. At present, most predictions of emission strengths rely on estimates using the quadrupole formula. This is justifiable because, given the uncertainties in our astrophysical understanding of potential sources, more accurate calculations would be unjustified in most cases. There are two exceptions to this generality. One is binary orbits, where the point-mass approximation is good over a large range of observable frequencies, so fully relativistic calculations (using the post-Newtonian methods described above) are not only possible, but are necessary for the construction of sensitive search templates in the data analysis. The second exception are the numerical simulations of the merger of black holes and neutron stars, where the dynamics is so complex that none of our analytic approximations offers us reliable guidance. 

When the interpretation phase of source calculations begins, one can anticipate that calculations will be very detailed, and strongly guided  by observed parameters (source masses, waveforms, and so on).  Their aim will be to model the observed signal so well that all of its details can be interpreted in terms of characteristics of the source: the total angular momentum of a system, the spins of individual stars, the equation of state of nuclear matter, and so on.

\begin{figure}
\epsfxsize=\textwidth 
\epsffile[31 220 567 597]{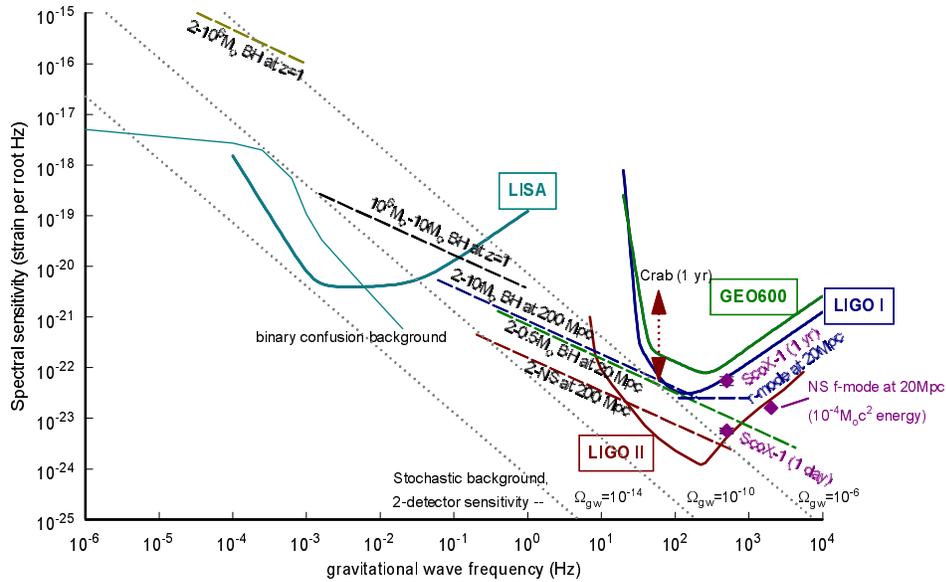}
\caption{Sensitivity of gravitational wave interferometers across 10 decades of frequency.  The solid curves show the expected and planned sensitivities of 4 representative interferometers. Other important instruments, such as VIRGO, have been omitted for the sake of clarity in the diagram, because their sensitivity lies between those of the instruments shown. The vertical axis is the strain spectral density, which is the square-root of the noise power per unit frequency.  The strength of the signals expected from coalescing systems are shown as spectral densities, in such as way that the area between a signal curve and an instrumental noise curve indicates the signal-to-noise ratio that could be achieved with perfect matched filtering; the signal spans the frequency range of the last 1 year before coalescence. Stochastic background sensitivities are shown as the limit that can be reached at 90\% confidence in any decade of frequency with an ``ideal'' correlation experiment between two co-located interferometers that have uncorrelated instrumental noise. Various possible neurton-star signals are shown, including the Crab pulsar, shown as a line from its upper limit (set by equating the observed loss of energy in spindown to the gravitational wave luminosity) to a point where its ellipticity would be $10^{-5}$. Two points are shown for hypothetical radiation that might come from Sco X-1.  The higher point is the spectral density when it is observed for a year, over which time modelling the signal might be difficult because fluctuations in accretion rate induce spin changes in the star.  The lower point requires only a 1-day observation, during which the signal should have a constant frequency. The predicted binary confusion limit noise at low frequencies is shown where it competes with the LISA sensitivity. }
\label{fig:sens}
\end{figure}

In this section I will review the major predictions that have beem made about likely gravitational wave sources, and I will examine the potential science that can be extracted by careful interpretation of the observations.  The discussions of this section are put into the context of the sensitivity of the detectors described earlier by Figure~\ref{fig:sens}, which is a composite overview of gravitational wave detection across 10 decades of frequency.

\subsubsection{Quadrupole approximation.}
The lowest order of the post-Newtonian approximation is the {\em quadrupole formula}, which gives the first approximation to the radiation emitted by a weakly relativistic system. If we define the spatial tensor $Q_{jk}$, the second moment of the mass distribution:
\begin{equation}\label{eqn:ibar}
Q_{jk}=\int\rho x_jx_k d^3x, 
\end{equation}
then  the amplitude of the gravitational wave (in Lorentz gauge but not TT gauge) is
\begin{equation}\label{eqn:hjk}
h_{jk} = \frac{2}{r}\oderivn{Q_{jk}}{t}{2}.
\end{equation}

To get the TT-amplitude of a wave travelling outwards from its source, project this tensor perpendicular to its direction of travel and remove the trace of the projected tensor. 

\subsubsection{Simple estimates.}
If the motion inside the source is highly non-spherical, then a typical component of $\oderivnf{Q_{jk}}{t}{2}$ will (from \Eqref{eqn:ibar}) have magnitude $Mv^2_{N.S.}$, where $v^2_{N.S.}$ is the non-spherical part of the squared velocity inside the source. So one way of approximating any component of \Eqref{eqn:hjk} is 
\begin{equation}\label{eqn:nonsph}
h\sim \frac{2Mv^2_{N.S.}}{r}.
\end{equation}
It is interesting to observe that the ratio $\epsilon$ of the wave amplitude to the Newtonian potential $\phi_{ext}$ of its source at the observer's distance $r$ is simply 
\[\epsilon \sim 2v^2_{N.S.}.\]
By the virial theorem for self-gravitating bodies, this will not be larger than 
\begin{equation}\label{eqn:epsilonandphi}
\epsilon <  \phi_{int},
\end{equation}
where $\phi_{int}$ is the maximum value of the Newtonian gravitational potential {\em inside} the system.  This provides a convenient bound in practice:
\begin{equation}
h < \phi_{int}\phi_{ext}.
\end{equation}

For a neutron star source one has $\phi_{int} \sim 0.2$.  If the star is in the Virgo cluster, then the upper limit on the amplitude of the radiation from such a source is $5\E{-22}$. This is a simple way to get the number that has been  the goal of detector development for decades, to make detectors that can observe waves at or below an amplitude of about $10^{-21}$.

\subsubsection{Estimating the frequency of the gravitational waves.}
The signals for which the best waveform predictions are available have narrowly defined frequencies. In some cases the frequency is dominated by an existing motion, such as the spin of a pulsar.  But in most cases the frequency will be related to the {\em natural frequency} for a self-gravitating body, defined as
\begin{equation}\label{eqn:natfreq}
f_0 = \sqrt{\bar{\rho}/4\pi},
\end{equation}
where $\bar{\rho}$ is the mean density of mass-energy in the source.  This is of the same order as the binary orbital frequency and the fundamental pulsation frequency of the body.  

The frequency is determined by the size $R$ and mass $M$ of the source, taking $\bar{\rho} = 3M/4\pi R^3$.  For a  neutron star of mass $1.4\msolar$ and radius 10~km, the natural frequency is $f_0 = 1.9$~kHz.  For a black hole of mass $10\msolar$ and radius $2M = 30$~km, it is $f_0 = 1$~kHz.  And for a large black hole of mass $2.5\E6\msolar$, such as the one at the center of our Galaxy, this goes down in inverse proportion to the mass to $f_0 = 4$~mHz.  

\begin{figure}
\epsffile[120 300 460 550]{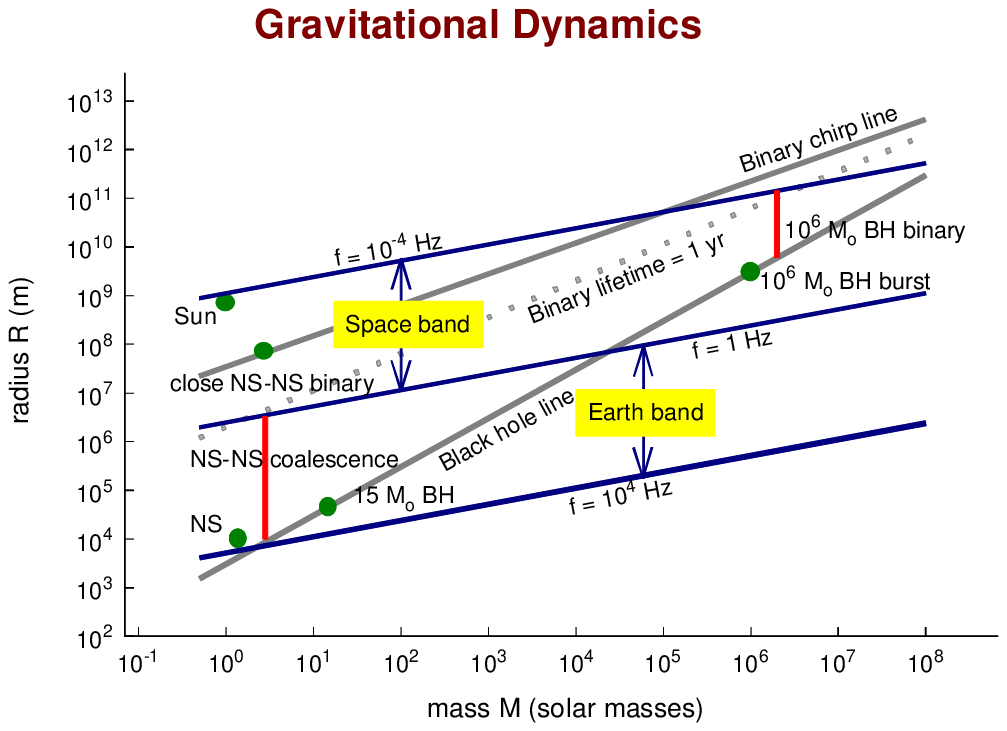}
\caption{Mass-radius plot for gravitational wave sources. }
\label{fig:freq}
\end{figure}

\Figref{fig:freq} shows the mass-radius diagram for likely sources of gravitational waves.  Three lines of constant natural frequency are plotted: $f_0 = 10^4$~Hz, $f_0 = 1$~Hz, and $f_0 = 10^{-4}$~Hz.  These are interesting frequencies from the point of view of observing techniques: gravitational waves between 1 and $10^4$~Hz are accessible to ground-based detectors, while lower frequencies are observable only from space. Also shown is the line marking the black-hole boundary.  This has the equation $R=2M$.  There are no objects below this line.  This line cuts through the ground-based frequency band in such a way as to restrict ground-based instruments to looking at stellar-mass objects.  Nothing over a mass of about $10^4\msolar$ can radiate above 1~Hz.  

A number of typical relativistic objects are placed in the diagram: a neutron star, a binary pair of neutron stars that spirals together as they orbit, some black holes.  Two other interesting lines are drawn.  The lower (dashed) line is the 1-year coalescence line, where the orbital shrinking timescale in \Eqref{eqn:chirp} is less than one year.  The upper (solid) line is the 1-year chirp line: if a binary lies below this line then its orbit will shrink enough to make its orbital frequency increase by a measurable amount in one year.  (In a one-year observation one can in principle measure changes in frequency of $1\units yr^{-1}$, or $3\E{-8}$~Hz.)  

It is clear from the figure that {\em any binary system that is observed from the ground will coalesce within an observing time of one year.}  Since pulsar statistics suggest that this happens less than once every $10^5$ years in our Galaxy, ground-based detectors must be able to register these events in a volume of space containing at least $10^6$ galaxies in order to have a hope of seeing occasional coalescences. 

\subsubsection{Polarization of gravitational waves.}
Gravitational waves have two independent polarisations, usually called $+$ and $\times$.\cite{MTW}  A general wave will have a linear combination of both. Rotating sources typically emit both polarisations with a phase delay between them, leading to elliptical polarisation patterns.  The polarisation of the waves contains important information. For example, a binary system emits purely circular polarisation along the angular momentum axis, but purely linear in its equatorial plane. By measuring the polarisation of waves from a binary (or from a spinning neutron star) one can determine the orientation and inclination of its spin axis.  This is a piece of information that is usually very hard to extract from optical observations.

\subsubsection{Luminosity in gravitational waves.}
The gravitational wave luminosity in the quadrupole approximation is
\begin{equation}\label{eqn:luminosity}
L_{gw} = \frac{1}{5}\left(\sum_{j,k}\qddd_{jk}\qddd_{jk} - \frac{1}{3}\qddd^2\right),
\end{equation}
where $Q$ is the trace of the matrix $Q_{jk}$. 
This equation can be used to estimate the back-reaction effect on a system that emits gravitational radiation. 

\subsection{Astronomical sources of gravitational waves}
Until observations of gravitational waves are successfully made, one can only make intelligent guesses about most of the sources that will be seen.  There are many that {\em could} be strong enough to be seen by the early detectors: binary stars, supernova explosions, neutron stars, the early Universe.  The detectability depends, of course, not only on the intrinsic luminosity of the source, but on how far away it is. Often the biggest uncertainty in making predictions is the spatial density of any particular class of sources. The estimates in this section use the quadrupole formula and other approximations, and so they are accurate only to within factors of order 2; but they show how important observables scale with the properties of the systems.

\subsubsection{Man-made gravitational waves.}
One source can unfortunately be ruled out as undetectable: man-made gravitational radiation.  Imagine creating a wave generator with the following extreme properties.  It consists of two masses of $10^3$~kg each (a small car) at opposite ends of a beam 10~m long.  At its center the beam pivots about an axis.  This centrifuge rotates 10 times per second. All the velocity is non-spherical, so $v^2_{N.S.}$ in \Eqref{eqn:nonsph} is about $10^5\units m^2\;s^{-2}$.  The frequency of the waves will actually be 20~Hz, since the mass distribution of the system is periodic with a period of 0.05~s, only half the rotation period.  The wavelength of the waves will therefore be $1.5\E7$~m, about the diameter of the earth.  In order to detect gravitational waves, not near-zone Newtonian gravity, the detector must be at least one wavelength from the source.  Then the amplitude $h$ can be deduced from \Eqref{eqn:nonsph}: $h \sim 5\E{-43}$.  This is far too small to contemplate detecting!

\subsubsection{Radiation from a spinning neutron star.}
Some likely gravitational wave sources behave like the centrifuge, however, only on a grander scale.  Suppose a  neutron star of radius $R$ spins with a frequency $f$ and has an irregularity, a bump of mass $m$ on its otherwise axially symmetric shape.  Then the bump will emit gravitational radiation (again at frequency $2f$ because it spins about its center of mass, so it actually has mass excesses on two sides of the star), and the non-spherical velocity will be just $v_{N.S.} = 2\pi Rf$.  The radiation amplitude will be, from \Eqref{eqn:nonsph}, 
\begin{equation}\label{eqn:bump}
h_{bump}\sim 2(2\pi Rf)^2m/r,
\end{equation}
and the luminosity, from \Eqref{eqn:luminosity} (assuming roughly 4 comparable components of $Q_{jk}$ contribute to the sum),
\[L_{bump}\sim (1/5)(2\pi f)^6m^2R^4.\]
The radiated energy would presumably come from the rotational energy of the star. This would lead to a spindown of the star on a timescale
\begin{equation}
t_{spindown} = \frac{1}{2}mv^2/L_{bump} 
\sim \frac{5}{4\pi}f^{-1}\fracparen{m}{R}^{-1}v^{-3}.
\end{equation}
It is believed that neutron star crusts are not strong enough to support asymmetries with a mass of more than about $m\sim10^{-5}\msolar$, and from this one can estimate the likelihood that the observed spindown timescales of pulsars are due to gravitational radiation.  In most cases, it seems that gravitational wave losses cannot be the main spindown mechanism.  

But lower levels of radiation would still be observable by detectors under construction, and this may be coming from a number of stars.  Particular interest has focussed recently on low-mass X-ray binary systems (LMXBs).  The neutron stars in them are accreting mass and angular momentum, so they should be spinning up.  As described earlier, observations suggest that most of the neutron stars are spinning at about 300~Hz.  Bildsten's proposed explanation\cite{Bildsten1998}  is still speculative, but the numbers make a plausible case that this radiation could carry away as much angular momentum as is accreted, leading to a limit on the spinup of such systems.  It is also possible that the r-mode instability mentioned earlier could operate in these stars: their temperature is marginal for reducing viscosity to a low enough level. In either case, the stars could turn out be long-lived sources of gravitational waves.  

\subsubsection{Radiation from a binary star system.}
Another ``centrifuge'' is a binary star system. Two stars of the same mass $M$ in a circular orbit of radius $R$ have $v^2_{N.S.} = M/4R$.  The gravitational-wave amplitude can then be written
\begin{equation}\label{eqn:binary}
h_{binary} \sim  \frac{1}{2}\frac{M}{r}\frac{M}{R}.
\end{equation}

The gravitational-wave luminosity of such a system is, by a calculation analogous to that for bumps on neutron stars, 
\[L_{binary} \sim \frac{1}{80G}\fracparen{M}{R}^5.\]
This equation is dimensionless, but can be converted to normal luminosity units by multiplying by the scale factor $c^5/G = 3.6\E{52}$~W.  By comparison, the luminosity of the Sun is only $3.8\E{26}$~W.  Close binaries can therefore radiate more energy in gravitational waves than in light. 

The radiation of energy by the orbital motion causes the orbit to shrink. The shrinking will make any observed gravitational waves increase in frequency
with time.  This is called a {\em chirp}. The timescale for this is
\begin{equation}\label{eqn:chirp}
t_{chirp} = Mv^2/L_{binary} \sim 20M\fracparen{M}{R}^{-4}.
\end{equation}

A more careful calculation shows that, for unequal masses, the quadrupole amplitude and the rate of shrinking depend on the masses only through the combination 
\begin{equation}
 \mptr = \mu^{3/5}M^{2/5},
\end{equation}
where $\mu$ is the reduced mass and $M$ the total mass. This is called the {\em chirp mass}.  If one can observe, in gravitational radiation, the shrinking time, 
then one can infer the chirp mass. If one then measures the amplitude of the radiation, the only unknown is the distance $r$ to the source. Gravitational wave observations of orbits that shrink because of gravitational energy losses can therefore directly determine the distance to the source. This is another way in which gravitational wave observations are complementary to electromagnetic ones, providing information that is hard to obtain electromagnetically.  One consequence is the possibility that observations of coalescing compact object binaries could allow one to measure the Hubble constant\cite{SCHUTZ1986} or other cosmological parameters. This will be particularly interesting for the LISA project, whose observations of black hole binaries could  contribute to the debate over the value of the cosmological constant.

The amplitude of the radiation from such a system is not a good guide to its detectability. As we will discuss below, data analysis techniques like matched filtering are able to eliminate most of the detector noise and allow the recognition of weaker signals. The improvement in amplitude sensitivity is roughly proportional to the square root of the number of cycles of the waveform that one observes.  For neutron stars that are observed from a frequency of 10~Hz until they coalesce, there could be of the order of $10^4$ cycles, meaning that the sensitivity of a second-generation detector would effectively be 100 times better than its broadband sensitivity.  Such detectors would see coalescences to 400~Mpc or more.  The event rate, determined from the small-number statistics of observed binary pulsar systems in our Galaxy, is estimated to be one event per year out to 200~Mpc,\cite{Lorimer} so second-generation detectors should see events once per month or so.  

As mentioned earlier, orbital shrinking has already been observed in the Hulse-Taylor Pulsar system, containing the radio pulsar PSR1913+16 and an unseen neutron star in a binary orbit. 
The key to the importance of this binary system is that all of the important parameters of the system can be measured before one takes account of the orbital shrinking.  This is because a number of post-Newtonian effects on the arrival time of pulses at the Earth, such as the precession of the position of the periastron and the time-dependent gravitational redshift of the pulsar period as it approaches and recedes from its companion, are measured and fully determine the masses and separation of the stars and the inclination and eccentricity of their orbit.  From these numbers, without any free parameters, it is possible to compute the shrinking timescale predicted by general relativity.  The observed rate matches the predicted rate to within the observational errors of less than 1\%.

Binary systems at lower frequencies are much more abundant than coalescing binaries and they have much longer lifetimes. LISA would look for close white-dwarf binaries in the Galaxy, and would probably see thousands of them.  For each resolved binary LISA could determine the orbital period and the spatial orientation of the orbit, and it could give a crude position.  If the orbit is seen to decay during the observation, LISA could determine the distance to the binary.  In fact, there are likely to be so many binaries that they will provide a confusion-dominated background radiation at frequencies below about 1~mHz, as shown in Figure~\ref{fig:sens}.  This is likely to be larger than instrumental noise at these frequencies, but it will not obscure strong black-hole signals.

\subsubsection{Neutron-star normal modes.}
In \Figref{fig:freq} there is a dot for the typical  neutron-star.  The corresponding frequency is the fundamental vibrational frequency of such an object.  In fact, neutron stars have a rich spectrum of non-radial normal modes, which fall into several families: f-, g-, p-, w-, and r-modes have all been studied.  If their gravitational wave emissions can be detected, then the details of their spectra would be a sensitive probe of their structure and of the equation of state of neutron stars, in much the same way that helioseismology probes the interior of the Sun.  This is a challenge to ground-based detectors, which cannot yet make sensitive observations as high as 10~kHz. It may be that specialised resonant-mass detectors or third-generation interferometers will open up this important part of the spectrum.

\subsubsection{Gravitational collapse.}
The event that forms a neutron star is the gravitational collapse that results in a supernova.  It is difficult to predict the waveform or amplitude expected from this event.  Although detecting this radiation has been a goal of detector development for decades, little more is known about what to expect than 30 years ago.  The burst might be broad-band, centred on 1~kHz, or it might be a few cycles of radiation at a frequency anywhere between 100~Hz and 10~kHz, chirping up or down.  If the emitted energy is more than about $0.01\msolar$, then second-generation detectors should have no trouble seeing events that occur in the Virgo Cluster.  But it is possible that amplitudes are much smaller. Axisymmetric simulations of gravitational collapse with neutrino transport show a kind of bubbling, or boiling, of the material outside the neutron star, as the neutrinos stream through it.\cite{BURROWS1995b} This material has relatively low density, and the frequencies are not high, but such boiling might be detectable by second-generation detectors from a supernova in the Galaxy.

If gravitational collapse forms a neutron star spinning rapidly, then it may be followed by a relatively long period (perhaps a year) of emission of nearly monochromatic gravitational radiation, as the r-mode instability described earlier forces the star to spin down to speeds of about 
100--200~Hz.\cite{Owen1999} If as few as 10\% of all the neutron stars formed since star formation began (at a redshift of perhaps 4) went through such a spindown, then they may have produced a detectable random background of gravitational radiation at frequencies down to 20~Hz.

\subsubsection{Gravitational waves from stellar-mass black holes.}
Astronomers now recognize that there is an abundance of black holes in the universe.  Observations across the electromagnetic spectrum have located black holes in X-ray binary systems in the Galaxy and in the centers of galaxies.  

These two classes of black holes have very different masses.  Stellar black holes typically have masses of around $10\msolar$, and are thought to have been formed by the gravitational collapse of the center of a large, evolved red giant star, perhaps in a supernova explosion.  Black holes in galactic centers seem to have masses between $10^6$ and $10^{10}\msolar$, but their history and method of formation are not yet well understood.

Both kinds of black hole can radiate gravitational waves.  According to \Figref{fig:freq}, stellar black hole radiation will be in the ground-based frequency range, while galactic holes are detectable only from space. The radiation from an excited black hole itself is strongly damped, lasting only a few cycles at the natural frequency (\Eqref{eqn:natfreq} with $R=2M$):
\[f_{BH}\sim 2800 \fracparen{M}{10\msolar}^{-1}\rm\; Hz.\]

Radiation from stellar-mass black holes is expected mainly from coalescing binary systems, when one or both of the components is a black hole.  Although black holes are formed more rarely than neutron stars, the spatial abundance of binary systems consisting of neutron stars with black holes, or of two black holes, may in fact be similar to the abundance of binary neutron stars. This is because binary systems are much more easily broken up when a neutron star forms than when a black hole forms. When a neutron star forms, most of the progenitor star's mass ($6\msolar$ or more) must be expelled from the system rapidly.  This unbinds the binary: the companion star has the same speed as before but is held to the neutron star by only  a  fraction of the original gravitational attraction.  When a black hole forms, most of the original mass may simply go down into the hole, and the binary will have a higher survival probability. 

Double black-hole binaries may in fact be ten times as abundant as neutron-star binaries, if recent suggestions are correct that globular clusters are efficient factories for black-hole binaries.\cite{Zwart} Being more massive than the average star in a globular cluster, black holes sink towards the centre, where three-body interactions can lead to the formation of binaries. The key point is that these binaries are not strongly bound to the cluster, so they can easily be expelled by later encounters.  From that point on they evolve in isolation, and typically have a lifetime shorter than $10^{10}$~y.  

The larger mass of such black hole systems makes them visible from a greater distance than neutron-star binaries. If binaries with black holes are this abundant, black hole events will be detected much more frequently than those involving neutron stars. They may even be seen by first-generation detectors like LIGO I and GEO600. By measuring the chirp mass (as discussed above) observers will recognize that they have a black-hole system.  It seems likely  that the first observations of binaries by interferometers will be of black holes.

More speculatively, black holes binaries may even be part of the dark matter of the universe.  Observations of MACHOs (microlensing of distant stars by compact objects in the halo of our galaxy) have indicated that up to half of the galactic halo could be in dark compact objects of $0.5\msolar$.\cite{MACHO,Sutherland1999} This is difficult to understand in terms of stellar evolution as we understand it today: neutron stars and black holes should be more massive than this, and white dwarfs of this mass should be bright enough to have been identified as the lensing objects. One possibility is that the objects were formed primordially, when conditions may have allowed black holes of this mass to form.  If so, there should also be a population of binaries among them, and occasional coalescences should therefore be expected.  In fact, the abundance would be so high that the coalescence rate might be as large as one every 20 years in each galaxy, which is higher than the supernova rate. Since binaries are maximally non-axisymmetric, these systems could be easily detected by first-generation interferometers out to the distance of the Virgo Cluster.\cite{Sasaki1997}

The estimates used here of detectability of black-hole systems rely entirely on the radiation emitted as the orbit decays, during which the point-particle post-Newtonian approximation should be adequate.  But the inspiral phase will of course be followed by a burst of gravitational radiation from the merger of the holes that will depend in detail on the masses and spins of the objects.  Numerical simulations of such events are needed to interpret this signal, and possibly even to extract it from the instrumental noise of the detector.  Moreover, for black holes of mass greater than about $50\msolar$, the inspiral phase ends at the frequency of the last stable orbit, which is perhaps 0.06 of the hole's natural frequency, or 30~Hz.  This is so near the lower limit of interferometer sensitivity that even detecting the event may require at least some information about the expected waveform.\cite{Flanagan1998} This aspect of gravitational wave detection may proceed at a pace determined by the success of numerical simulations of black-hole mergers.

\subsubsection{Massive and supermassive black holes.}
Gravitational radiation is expected from galactic-mass black holes in two ways.  In one scenario, two massive black holes spiral together in a much more powerful version of the coalescence we have just discussed.  The frequency is much lower, in inverse proportion to their masses, but the amplitude is higher.  \Eqref{eqn:chirpeffective} below implies that the effective signal amplitude is almost linear in the masses of the holes, so that a signal from two $10^6\msolar$ black holes will have an amplitude $10^5$ times bigger than the signal from two $10\msolar$ holes at the same distance.  Even allowing for differences in technology, this indicates why space-based detectors will be able to study such events with a very high signal-to-noise ratio no matter where in the universe they occur.  
Observations of coalescing massive black-hole binaries will therefore provide unique insight into the behaviour of strong gravitational fields in general relativity. 

The event rate for such coalescences is not easy to predict,  but is likely to be large.  It seems that the central core of most galaxies may contain a black hole of at least $10^6\msolar$.  This is known to be true for our galaxy\cite{ECKART1996} and for a very large proportion of other galaxies that are near enough to be studied in sufficient detail.\cite{Richstone}  Supermassive black holes (up to a few times $10^9\msolar$) are believed to power quasars and active galaxies, and there is some evidence that the mass of the central black hole is proportional to the mass of the core of the host galaxy.\cite{Richstone2}  

If black holes are formed with their galaxies, in a single spherical gravitational collapse event, and if nothing happens to them after that, then coalescences will never be seen.  But this is unlikely for two reasons. First, it is believed that galaxy may have occurred through the merger of smaller units, sub-galaxies of masses upwards of $10^6\msolar$.  If these units had their own black holes, then the mergers would have resulted in the coalescence of many of the holes on a timescale shorter than the present age of the universe.  This would give an event rate of several mergers per year in the Universe, most of which would be observable by LISA if the more massive hole  is not larger than about $10^7\msolar$.  If the $10^6\msolar$ black holes were formed from smaller holes in a hierarchical merger scenario, then the event rate could be hundreds or thousands per year.  The second reason is that we see large galaxies merging frequently.  Interacting galaxies are common, and if galaxies come together in such a way that their central black holes both remain in the central core, then dynamical friction with other stars will bring them close enough together to allow gravitational radiation to bring about a merger on a timescale of less than $10^{10}$~y.\cite{HAEHNELT1993}

Besides mergers of holes with comparable masses, the capture of a small compact object by a massive black hole canalso result in observable radiation. The tidal disruption of main-sequence or giant stars that stray too close to the hole is thought to provide the gas that powers the quasar phenomenon.  These clusters will also contain a good number of neutron stars and stellar-mass black holes.  They are too compact to be disrupted by the hole even if they fall directly into it.  

Such captures therefore emit a gravitational wave signal that will be well approximated as that from a point mass near the black hole.  This will again be a chirp of radiation, but in this case the orbit may be highly eccentric.  The details of the waveform encode information about the geometry of space-time near the hole.  In particular, it may be possible to measure the mass and spin of the hole and thereby to test the uniqueness theorem for black holes.  The event rate is not very dependent on the details of galaxy formation, and is probably high enough for many detections per year from a space-based detector, provided that theoretical calculations provide data analysts with accurate predictions of the motion of these point particles over many tens of orbits.

Predicting these orbits may not be easy, however.  One problem that is not yet solved is to find a practical way of calculating the effects of radiation reaction on point particles, which we discussed earlier.  Another potential problem is that the motion of a particle on a highly eccentric orbit near a Kerr black hole may be effectively chaotic.  The orbit of such a particle --- particularly if it has spin, which is likely --- if it moves well outside of the equatorial plane is very complex,  and it probably wanders rapidly around the entire horizon.  In order to predict the motion for tens of orbits, it may be necessary to define the initial data, or other parameters that distinguish one motion from another, with a very high accuracy.  This would mean that one has to use a huge family of templates in order to pick out the signal (see our discussion of data analysis below).  In turn, this raises the confidence threshold necessary to eliminate false alarms in the detection process.  It is not yet clear whether this threshold will still allow signals of the expected strength to be detected.

\subsubsection{Gravitational waves from the Big Bang.}
Gravitational waves have traveled almost unimpeded through the universe since they were generated.  Whereas the  cosmic microwave background reflects the universe at a time $10^5$~y after the Big Bang, and studies of nucleosynthesis reveal conditions in the universe a few minutes after the Big Bang, gravitational waves were produced at times earlier than $10^{-24}$~s after the Big Bang.  Observing this background would undoubtedly be one of the most important measurements that gravitational wave astronomy could make.

Inflation is an attractive scenario for the early universe because it makes the large-scale homogeneity of the universe easier to understand.  It also provides a mechanism for producing initial density perturbations large enough to evolve into galaxies as the universe expands.  These perturbations are accompanied by gravitational-field perturbations that travel through the universe, redshifting in the same way that photons do.  Today these perturbations should form a random background of gravitational radiation.\cite{ALLEN1988}

The perturbations arise by parametric amplification of quantum fluctuations in the gravitational wave field that existed before inflation began.\cite{GRISHCHUK1975}  The huge expansion associated with inflation puts energy into these fluctuations, converting them into real gravitational waves with classical amplitudes. 

If inflation did not occur, then the perturbations that led to galaxies must have arisen in some other way, and it is possible that this alternative mechanism also produced gravitational waves.  One candidate is cosmic defects, including {\em cosmic strings} and cosmic texture.  Although observations at present seem to rule cosmic defects out as a candidate for galaxy formation, cosmic strings may nevertheless have produced observable gravitational waves.\cite{CALDWELL1996}

There could also be a thermal background under certain circumstances.  If inflation did not occur, but at the Planck time there was some kind of equipartition between gravitational degrees of freedom and other fields, then there would also be a thermal background of gravitational waves at a temperature similar to that of the cosmological microwave background.  But this radiation would have such a high frequency that it would not be detectable by any known or proposed technique.  If inflation occurred, it would have redshifted this background down to undetectable frequencies.

The random background will be detectable as a noise, competing with instrumental noise.  To be observed by a single detector, this noise must be larger than the instrumental noise, and one must have great confidence in the detector in order to claim that the observed noise is external.  This is how the cosmic microwave background was originally discovered in a radio telescope.  It is how LISA would look for a background, since the two interferometers in LISA would share a common arm and would therefore not be able to improve their sensitivity by cross-correlation, as described next.

If there are two detectors, then one can cross-correlate their output.  In this way, the random wave field in one detector acts like a matched filtering template, matching the random field in the other detector.  This allows the detection of noise that is below the instrumental noise of the individual detectors.  For this to work, the two detectors must be close enough together to respond to the same random wave field.  In practice, the sensitivity of this method falls off rapidly with separation if the detectors are more than a wavelength apart.  The two LIGO detectors are well-placed for such correlations, particularly when upgrades push their lower frequency limit to 20~Hz or less.

Random gravitational waves are conventionally described in terms of their energy density $\rho_{gw}(f)$ rather than their mean amplitude. For a cosmological field, what is relevant is to normalize this energy density to the critical density $\rho_c$ required to close the universe.  It is thus conventional to define 
\begin{equation}\label{eqn:cosdef}
\Omega_{gw}:= \oderiv{\rho_{gw}/\rho_c}{\ln f}.
\end{equation}
This is roughly the fraction of the closure energy density in random gravitational waves between the frequency $f$ and $e\times f$.  

Inflation predicts that the $\Omega_{gw}$ spectrum should  be flat, independent of frequency.  Other models, such as string cosmologies, can make very different predictions,\cite{BRUSTEIN1995} so it is possible that there will be a peak in the spectrum in the observing band of ground-based or space-based detectors.  

The first-generation LIGO detectors should reach a sensitivity of $\Omega_{gw} \sim 10^{-5}$ at 100~Hz, and second-generation LIGO could reach $10^{-9}$ at a similar frequency.  LISA could also approach $10^{-8}$ at about 1~mHz, but the exact sensitivity will depend on whether a binary background obscures the cosmological background. There is also a likelihood that backgrounds due to other sources (r-mode spindown, as discussed above, or binary black holes and neutron star systems, or even supernova explosions themselves\cite{Ferrari}) could obscure a cosmological background above 0.1~mHz.  In fact, it is likely that the  binary star background falls off (as measured by $\Omega_{gw}$) at frequencies below about $10\;\mu$Hz, so that a space-based detector designed for the microHertz frequency band might be the best way to observe the background.

\subsection{Recognizing weak signals}\label{sec:data}
For ground-based detectors, all expected signals have amplitudes that are close to or even below the instrumental noise level in the detector output.  Such signals can nevertheless be detected with confidence if their waveform matches an expected waveform.  The pattern recognition technique that will be used by detector scientists is called {\em matched filtering}.  

Matched filtering works by multiplying the output of the detector by a function of time (called the {\em template}) that represents an expected waveform, and summing (integrating) the result.  If there is a signal matching the waveform buried in the noise then the output of the filter will be higher than expected for pure noise. 

A simple example of such a filter is the Fourier transform, which is a matched filter for a constant-frequency signal.  The noise power in the data stream is spread out over the spectrum, while the power in the signal is concentrated in a single frequency.  This makes the signal easier to recognize.  The improvement of the signal-to-noise ratio for the amplitude of the signal is proportional to the square-root of the number of cycles of the wave contained in the data.  This is well-known for the Fourier transform, and it is generally true for matched filtering.  

Matched filtering can make big demands on computation, for several reasons. First, the arrival time of a short-duration signal is generally not known, so the template has to be multiplied into the data stream at each distinguishable arrival time.  This is then a correlation of the template with the data stream. Normally this is done efficiently using Fast Fourier Transform methods.  

Second, the expected signal usually depends on a number of unknown parameters.  For example, the radiation from a binary system depends on the chirp mass $\mptr$, and it might arrive with an arbitrary phase.  Therefore, many related templates must be separately applied to the data to cover the whole family.

Third, matched filtering enhances the signal only if the template stays in phase with the signal for the whole data set.  If they go out of phase, the method begins to reduce the signal-to-noise ratio.  For long-duration signals, such as for low-mass neutron-star coalescing binaries or continuous-wave signals from neutron stars (see below), this requires the analysis of large data sets, and often forces the introduction of additional parameters to allow for small effects that can make the signal drift out of phase.  It also means that the method works well only if there is a good prediction of the form of the signal.  

Because the first signals will be weak, matched filtering will be used wherever possible.  As a simple rule of thumb, the detectability of a signal depends on its {\em effective} amplitude $h_{eff}$, defined as 
\begin{equation}\label{eqn:effective}
h_{eff} = h \sqrt{N_{cycles}},
\end{equation}
where $N_{cycles}$ is the number of cycles in the waveform that are matched by the template.  

For example, the effective amplitude of the radiation from a bump on a neutron star (\Eqref{eqn:bump})will be $h_{bump}\sqrt{2fT_{obs}}$, where $T_{obs}$ is the observation time.  In order to detect this radiation, detectors may need to observe for long periods, say 4 months, during which they accumulate billions of cycles of the waveform.  During this time, the star may spin down by a detectable amount, and the motion of the Earth introduces large changes in the apparent frequency of the signal, so matched filtering needs to be done with care and precision. In order to do a sky survey for unknown neutron star sources, the data analysis job is so large that the sensitivity of detectors may be limited by the computer power available to process the data.\cite{SCHUTZ1991a,Brady1998}  

Another example is a binary system followed to coalescence, {\em i.e.\ } where the chirp time in \Eqref{eqn:chirp} is less than the observing time.  For neutron-star binaries observed by ground-based detectors this will always be the case (see above), so the effective amplitude is roughly 
\begin{equation}\label{eqn:chirpeffective}
h_{chirp} \sim h_{bin}\sqrt{f_{gw}t_{chirp}} \sim  \frac{GM}{rc^2}\fracparen{GM}{Rc^2}^{-1/4},
\end{equation}
where for $f_{gw}$ one must use twice the orbital frequency $\sqrt{GM/R^3}/4\pi$.  This may seem a puzzling result, because it says that the effective amplitude of the signal gets {\em smaller} as the stars get closer.  But this just means that the signal will be more detectable if it is picked up earlier, since \Eqref{eqn:chirpeffective} assumes that the signal is followed right to coalescence.  This gives a significant advantage to detectors that can operate at lower frequencies.

In general, the sensitivity of detectors will be limited not just by detector technology, but also by the duration of the observation, the quality of the signal predictions, and the availability of computer processing power for the data analysis.

\section{Conclusions}
The first few years of the 21st century should see the first direct detections of gravitational radiation and the opening of the field of gravitational wave astronomy.  Beyond that, over a period of a decade or more, one may expect observations to yield important and useful information about binary systems, stellar evolution, neutron stars, black holes, strong gravitational fields, and cosmology.

If gravitational wave astronomy follows the example of other fields, like  X-ray astronomy and radio astronomy, then at some level of sensitivity it will begin to discover sources that were completely unexpected.  Many scientists think the chance of this happening early is very good, since the processes that produce gravitational waves are so different from those that produce the electromagnetic radiation on which most present knowledge of the universe is based, and since more than 90\% of the matter in the universe today is dark and seems to interact with visible matter only through gravitation.  By looking at the dynamical part of the gravitational field, gravitational wave detectors are the only way of discovering whether interesting things are happening in this dark sector of the Universe.

The first and second generation of ground-based interferometers, and the first space-based project, are all very likely to operate within the first decade of the 21st century. They will open up the observational side of gravitational wave astronomy.  But we know that, sophisticated as their designs are, they are not ideal for some kinds of gravitational wave sources that we believe exist.  Sensitive measurements of a cosmological background of radiation from the big bang may not be possible with these instruments if the spectrum follows the predictions of standard inflation theory.  Most of the normal mode oscillations of neutron stars will be very hard to detect, because the radiation is weak and at a high frequency, but the science there is compelling: neutron-star seismology may be the only way to probe the interiors of neutron stars and understand these complex and fascinating objects.  The field of gravitational wave astronomy can therefore be expected to have a long future.  There will be much to do for a long time after the first observations are successfully made.

\end{document}